\documentclass[aps,prl,reprint,twocolumn,superscriptaddress]{revtex4-1}

\usepackage[dvips]{graphicx,color}
\usepackage{amsmath,amssymb}
\usepackage{bm}

\begin{document}


\title{Observation of the Spatial Distribution of Gravitationally Bound Quantum States
of Ultracold Neutrons and its Derivation Using the Wigner Function}

\author{G.~Ichikawa}
\author{S.~Komamiya}
\author{Y.~Kamiya}
\author{Y.~Minami}
\author{M.~Tani}
\affiliation{Department of Physics, Graduate School of Science, and International Center
for Elementary Particle Physics, The University of Tokyo, 7-3-1 Hongo, Bunkyo-ku,
Tokyo 113-0033, Japan}

\author{P.~Geltenbort}
\affiliation{Institut Laue-Langevin, BP 156, 6, rue Jules Horowitz,
38042 Grenoble Cedex 9, France}

\author{K.~Yamamura}
\author{M.~Nagano}
\affiliation{Research Center for Ultra-Precision Science and Technology,
Graduate School of Engineering, Osaka University, 2-1 Yamadaoka, Suita,
Osaka 565-0871, Japan}

\author{T.~Sanuki}
\affiliation{Department of Physics, Graduate School of Science, Tohoku University,
6-3, Aramaki Aza-Aoba, Aoba-ku, Sendai 980-8578, Japan}

\author{S.~Kawasaki}
\affiliation{High Energy Accelerator Research Organization,
Institute of Particle and Nuclear Studies, 1-1 Oho, Tsukuba, Ibaraki 305-0801, Japan}

\author{M.~Hino}
\affiliation{Kyoto University Research Reactor Institute, 2, Asahiro-Nishi,
Kumatori-cho, Sennan-gun, Osaka 590-0494, Japan}

\author{M.~Kitaguchi}
\affiliation{Department of Physics, Graduate School of Science, Nagoya University,
Furo-cho, Chikusa-ku, Nagoya 464-8601, Japan}

\date{\today}

\begin{abstract}
  Ultracold neutrons (UCNs) can be bound by the potential of terrestrial gravity and
  a reflecting mirror.
  The wave function of the bound state has characteristic modulations.
  We carried out an experiment to observe the vertical distribution of the UCNs
  above such a mirror at Institut Laue-Langevin in 2011.
  The observed modulation is in good agreement with that prediction by
  quantum mechanics using the Wigner function.
  The spatial resolution of the detector system is estimated to be $0.7\;\mu\mathrm{m}$.
  This is the first observation of gravitationally bound states of UCNs with
  submicron spatial resolution.
\end{abstract}


\maketitle 

Terrestrial gravity is the most common force experienced in everyday life.
However, experimental measurements of quantum mechanical bound states in the Earth's
gravitational field were started only in the last decade by the pioneering works of
Nesvizhevsky \textit{et al.\ } 
using ultracold neutrons (UCNs) \cite{Nesv_Nature, Nesv_EPJC}.
UCNs are neutrons with kinetic energies lower than the Fermi pseudo-potential of
materials (e.g. Ni, with $\sim 200\;\mathrm{neV}$)
and are hence totally reflected by the material surfaces at any angle of incidence.
The wave function $\psi(z)$ of a UCN in the terrestrial gravitational field obeys the
Schr\"{o}dinger equation in the vertical direction $z$.
The eigenstates of this system
are linear combinations of Airy functions \cite{Airy}.
The vertical probability distribution of UCN bound states,
namely, the sum of the absolute squares of eigenfunctions,
has a characteristic modulation.
The eigenstate is specified by two scales,
the length $\left(\hbar^2/\left(2m^2 g \right)\right)^{1/3} = 5.87\;\mu\mathrm{m}$
and the energy $ \left(mg^2 \hbar^2/2\right)^{1/3} = 0.602\;\mathrm{peV}$
where $\hbar$ is the reduced Planck constant, $m$ is the neutron mass
and $g$ is the gravitational acceleration.
There are therefore two ways to observe the bound state,
to measure its energy or its position.
Recently first measurements of the differences between eigenenergies using
the transitions of UCNs between quantum states in the terrestrial gravitational potential
have been reported \cite{qBounce}.
The ability of experiments to observe the spatial distribution of the bound state is
limited by the spatial resolution (about two microns) of current slow neutron detectors.

\begin{figure}[tbp!]
  \includegraphics[width=85mm]{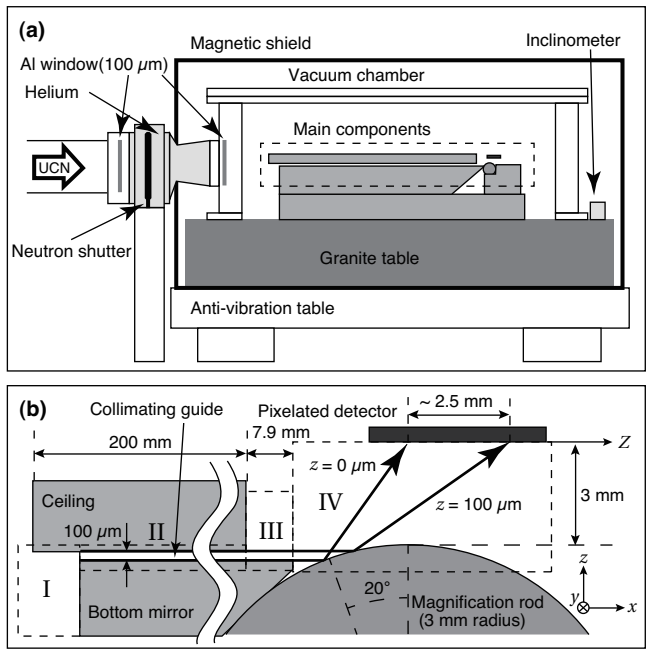}
  \caption{Experimental setup, (a) general view and (b) the main components.
  The two thick bent arrows in (b) correspond to the trajectories of neutrons flying
  horizontally above the bottom mirror with the height of the bottom mirror
  $z = 0\;\mu\mathrm{m}$ and that of the ceiling $z = 100\;\mu\mathrm{m}$.
  The $45^\circ$ slope of the bottom mirror and the rod are designed to come in
  contact with each other.
  The Roman numerals denote the calculation steps.
  \label{fig:setup}}
\end{figure}

To observe the spatial distribution of gravitationally bound states with high precision,
we developed a novel technique, shown in Figure \ref{fig:setup},
with three main components \cite{Sanuki}.
The Cartesian coordinate system $(x,y,z)$ is defined in Figure \ref{fig:setup} (b).
Incident UCNs pass through the collimating guide
in which they settle into gravitationally bound states above the flat bottom mirror.
The ceiling removes UCNs whose wave functions significantly penetrate the ceiling.
The height distribution of the surviving UCNs is magnified by a cylindrical rod
which acts as a convex mirror.
After reflection at the rod surface,
UCNs are detected by a CCD-based pixelated detector.
$Z$ is the axis on the pixelated detector corresponding to the magnified height $z$.
The collimating guide and magnification rod have a width of $50\;\mathrm{mm}$ along $y$.
Using this setup,
we performed an experiment to observe the spatial distribution of gravitationally
bound states during a period of 17 days in August 2011 at the Institut
Laue-Langevin (ILL).

We used the UCN beam line PF2 \cite{ILL} at ILL, the world's highest intensity
steady UCN source.
The horizontal velocity ($v_x$) distribution of UCNs was measured using
standard time-of-flight technique.
The measured velocity distribution is nearly Gaussian, with a mean of $9.4\;\mathrm{m}/\mathrm{s}$
and standard deviation of $2.8\;\mathrm{m}/\mathrm{s}$.

The energy of UCNs is quantized inside the collimating guide made of glass,
with a height $h=100\;\mu\mathrm{m}$.
The ceiling of the guide removes UCNs with high vertical energy due to its microscopic
surface roughness, with an arithmetic mean of $0.4\;\mu\mathrm{m}$.
Once such a neutron is reflected by the ceiling,
the large horizontal velocity component is converted into vertical velocity component.
The chance to hit the ceiling again is then enormously enhanced.
Numerous collisions cause UCN loss by absorption or upscattering.
In addition an absorptive Gd-Ti-Zr alloy ($54/35/11$)
was deposited on the glass by vacuum evaporation \cite{KUR} at the
Kyoto University Research Reactor Institute (KURRI).
The thickness of the layer is $200\;\mathrm{nm}$ and its potential is calculated to be
$-13.9-26.5 i\;\mathrm{neV}$.

At the end of the collimating guide, a cylindrical glass rod of radius $3\;\mathrm{mm}$
magnifies the distribution of UCN like a convex mirror.
The geometrical arrangement of the magnification system is shown in Figure
\ref{fig:setup} (b).
The distribution of $100\;\mu\mathrm{m}$ in height $z$ is magnified to
$\sim 2.5\;\mathrm{mm}$ in the position $Z$ on the detector,
hence the average magnification power is about 25.
The glancing angle was only $20^\circ$
in order to make the critical energy of reflection high.
Differences in $v_x$ cause dispersion of the parabolic trajectories
and smear the distribution.
This dispersion is estimated to be less than $0.1\;\mu\mathrm{m}$ of the height $z$.
The rod was precisely ground by Crystal Optics Inc. and
finely polished at the Research Center for Ultra-Precision Science and Technology,
Osaka University.
Furthermore a Ni layer of $200\;\mathrm{nm}$ was deposited on the polished glass surface,
increasing the potential from $100\;\mathrm{neV}$ to
$200\;\mathrm{neV}$ so that all UCNs exiting the guide were totally reflected.
The Ni deposited surface has an arithmetic mean roughness
of $1.9\;\mathrm{nm}$,
much smaller than the wavelength of UCN $\sim 100\;\mathrm{nm}$.
Hence, the diffused reflection from the surface roughness of the rod can
be neglected.

For high resolution two-dimensional detection, a charge-coupled device (CCD) was used.
Since slow neutrons would pass through the sensitive volume of CCDs
without ionization,
they must first be converted to charged particles.
To retain the intrinsic spatial resolution,
a $200\;\mathrm{nm}$ thin $^{10}\mathrm{B}$ neutron converter was evaporated
directly onto a CCD.
After neutron capture $\alpha$ and $^{7}\mathrm{Li}$ particles are released.
A back-thinned CCD \cite{Hamamatsu} was the base of the detector,
with a pixel size is $24\;\mu\mathrm{m} \times 24\;\mu\mathrm{m}$ and
sensitive area of $24.576~\mathrm{mm} \times 6.000~\mathrm{mm}$.
The length along the $Z$-axis is $6~\mathrm{mm}$.
An incident charged particle creates electron-hole pairs inside the Si layer and
loses energy.
About 300,000 electrons per $\mathrm{MeV}$ are created and spread before reaching
electrodes and being detected as a two-dimensional cluster.
The barycenter of the deposited charges corresponds to the incident position of
the charged particle.
The spatial resolution along the $Z$-axis was measured to be
$3.35\pm0.09\;\mu\mathrm{m}$ \cite{Kawasaki}.

The setup was installed inside a vacuum chamber to prevent
neutrons from interacting with air.
We evacuated the chamber to $10\;\mathrm{Pa}$ before the experimental run.
The vacuum pump was disconnected during the measurement to reduce vibration.

A neutron shutter with Cd blades was installed inside an acrylic box
as shown in Figure \ref{fig:setup} (a) to shut off the UCN beam during
the readout of the CCD.
The box was connected to the beam pipe and the vacuum chamber by plastic bellows,
to prevent the transmission of vibrations.
The shutter box and the plastic bellows were filled with Helium gas
to minimize scattering.

The magnetic shield of mu-metal covering the experimental apparatus reduces
the external magnetic field by about $1/100$.
As shown in Figure \ref{fig:setup} (a),
the detector system was installed on granite and anti-vibration tables
\cite{HERZ} to reduce vibration from the floor.

The horizontality of the detector system was better than $0.1\;\mathrm{mrad}$
and monitored by an inclinometer.
The remaining effects due to external magnetic field,
vibration and horizontality were estimated to be negligible.

The distribution of UCNs on the pixelated detector can be calculated using
quantum mechanics.
The state of a UCN can in general be written as a superposition of the $n$-th
energy eigenstates as
$\Psi(z,t) = \sum_n a_n \psi_n(z)\exp(- i E_n t/\hbar)$
where $a_n$ satisfies $\sum_n |a_n|^2=1$ and $\psi_n(z)$ is normalized to give
$\int d z\; |\psi_n(z)|^2=1$.
The experimental result is an average over many incoherent UCNs.
Since the phase of $a_n$ is randomly and uniformly distributed,
the average of the absolute squared value of the superposition becomes
$|\Psi(z)|^2 = \sum_n |a_n|^2 |\psi_n(z)|^2$
where the time dependence and interference terms are averaged out.
In general, the quantum state of UCNs is treated as a mixed state.
The state is described as a density matrix,
$\Hat{\rho} = \sum_n p_n |\psi_n\rangle \langle\psi_n|$
where $p_n = |a_n|^2$ is the probability of the $n$-th state and $|\psi_n\rangle$ is
the corresponding state vector.
The calculation was performed in four steps (I, II, III, IV) from upstream towards the detector,
as shown in Figure \ref{fig:setup} (b).
In the first three steps, the probability of the eigenstate $p_n$ at the end of
the guide is calculated and then, in the last step,
the UCN distribution on the pixelated detector is derived.
Each step is discussed in the following paragraphs.

(I) The probabilities of the eigenstates just after the entrance of the collimating guide
are assumed to be uniform.
It should be noted that the eigenstates with vertical energy larger than $mgh$
($h=100\;\mu\mathrm{m}$)
inside the guide are different from the case without the ceiling,
because the wave function becomes zero at the height of the ceiling.
In the following, values with tildes denote the case with the ceiling
at $z=h$.

(II) For UCN loss inside the guide, we assume a phenomenological loss rate of
the $n$-th state as $\Gamma_n + \mathrm{B}_n$ where $\Gamma_n$ and $\mathrm{B}_n$ are
respectively the loss by the ceiling and the bottom mirror.
$\Gamma_n$ is assumed to be proportional to the probability of finding a UCN in
the roughness region as
$
\Gamma_n
= \gamma \cdot \int_{h-2\delta}^{h}d z\;
| \Tilde{\psi}_n (z) |^2
$
where $\gamma$ is the constant for the loss and $\delta = 0.4\;\mu\mathrm{m}$ is
the arithmetic mean roughness of the ceiling
($2\delta$ is the average width of the roughness region) \cite{Westphal}.
$\mathrm{B}_n$ is assumed to be proportional to the bouncing number per unit time of
a bouncing motion above a floor in classical mechanics,
$
\mathrm{B}_n = \beta \cdot (g/2\sqrt{2}) \sqrt{m/\Tilde{E}_n},
$
where $\beta$ is the constant for the loss.
Hence, the probability at the guide exit $\Tilde{p}_n$ is written in terms of
that at the guide entrance $\Tilde{p}_n (0)$ as
$
\Tilde{p}_n \propto \Tilde{p}_n (0)
\left\langle \exp \left( - l/v_x\cdot (\Gamma_n+\mathrm{B}_n)\right)
\right\rangle
_{v_x}
$
where $l$ is the length of the guide, $\langle \; \rangle_{v_x}$
indicates the average over the measured distribution of $v_x$ and
the normalization constant is chosen to satisfy $\sum_n \Tilde{p}_n = 1$.

\begin{figure}[tbp!]
  \includegraphics[width=85mm]{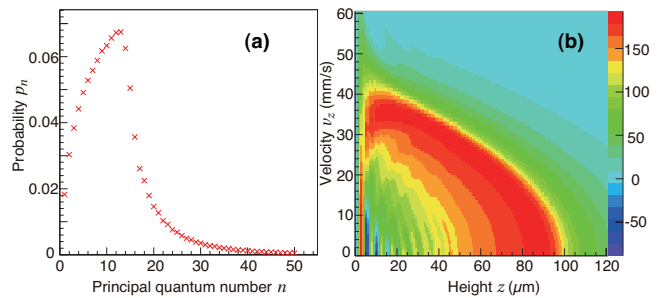}
  \caption{(color). The resulting probabilities for eigenstates $p_n$ (a) and
  the corresponding Wigner function $W(z,v_z)$, the sum of the Wigner functions
  $n\leq 50$ with the weights of the probabilities for the eigenstates (b).
  $W(z,v_z)$ for $v_z>0$ is shown here because $W(z,v_z)=W(z,-v_z)$.
  The color scale in (b) is in arbitrary units.
  The best fit parameters are used in both figures.
  \label{fig:pop_Wigner}}
\end{figure}

(III) At the exit of the guide, the wave function changes
because the ceiling suddenly disappears.
From the sudden approximation,
the wave function is continuous at the time of the sudden change as
$\sum_m \Tilde{a}_m \Tilde{\psi}_m = \sum_n a_n \psi_n$
where the left (right) hand side corresponds to the wave function with (without)
the ceiling.
Using the properties of a complete orthonormal system, we obtain
$p_n = \sum_m \Tilde{p}_m
\left|\int_0^{h}d z\; \psi_n(z)\Tilde{\psi}_m(z) \right|^2$
where the probabilities satisfy $p_n = |a_n|^2$ and $\Tilde{p}_m = |\Tilde{a}_m|^2$.
In this equation, the cross terms are cancelled due to random phases.
The resulting probabilities of the eigenstates $p_n$ are shown in
Figure \ref{fig:pop_Wigner} (a).
The suppression of small $n$ eigenstates has been reported by past experiments
\cite{Nesv_EPJC, Westphal}.

(IV) The detected position on the detector after magnification depends not only
on the height $z$ but also on the vertical velocity $v_z$ at the end of the guide.
Hence we use a kind of probability density in phase space given by the Wigner function
\cite{Wigner}.
The Wigner function is defined as
\[
   W(z,p_z) = \frac{1}{2\pi \hbar}
        \int_{-\infty}^{\infty} \!\!\!\! d \xi\;
        \psi^*(z-\tfrac{1}{2}\xi)
        \psi(z+\tfrac{1}{2}\xi)
        \exp\left(-\frac{i p_z\xi}{\hbar}\right)
        \label{eq:Wigner}
\]
where $z$ is the height, $p_z$ is the momentum along $z$ and
$\psi$ is the wave function.
The Wigner function in $z$-$v_z$ phase space can be obtained by replacing
$p_z$ with $m v_z$.
The Wigner function has properties of
$\int_{-\infty}^{\infty}d p_z\;W(z,p_z)=|\psi(z)|^2$
and
$\int_{-\infty}^{\infty}d z\;W(z,p_z)=|\varphi(p_z)|^2$
where $\varphi(p_z)$ is the momentum space wave function.
The Wigner function of a mixed state
$\Hat{\rho}=\sum_n p_n |\psi_n\rangle \langle \psi_n|$
is a simple sum 
$W(z,p_z) = \sum_n p_n W_n(z,p_z)$
where $W_n$ is the Wigner function for the $n$-th eigenstate.
The Wigner function obtained for this system is shown in
Figure \ref{fig:pop_Wigner} (b).
We calculate the correspondence of the phase space point at the guide end and
the detection point on the detector by the classical trajectory.
This treatment is supported by the fact that
if the potential of the system has terms only up to the second order in the position,
the motion of the Wigner function in the phase space can be obtained by the equation
of motion in classical mechanics \cite{Schleich}.
In this case, the potential has only first order term of $mgz$.
The whole phase space was divided into a mesh with a size of
$\Delta z \times \Delta v_z = 0.1\;\mu\mathrm{m} \times 0.1\;\mathrm{mm}/\mathrm{s}$
and each mesh point was weighted by the Wigner function $W(z,v_z)$.

The predicted distribution was fitted to the data using a binned maximum likelihood
method.
Six parameters were used in the fit, $\theta$, $Z_0$, $d$, $\gamma$, $\beta$ and $s$.
$\theta$ denotes the rotation of the pixelated detector in the detector plane.
The position of the data $Z$ is rotated by $\theta$ to be $Z \rightarrow Z'$.
$Z_0$ is the offset for the position of predicted events $Z_{\mathrm{pred}}$.
The relation of the coordinates is $Z'= Z_{\mathrm{pred}}+Z_0$.
$d$ is the difference between the actual and design heights of the pixelated detector,
hence $d$ effectively modifies the magnification power.
$d>0$ denotes that the actual position is higher than the design,
corresponding to a higher magnification.
$\gamma$ and $\beta$ are the parameters describing losses inside the guide.
$s$ is the ratio of the signal to the total events.
A flat background in $Z$ is assumed.
The predicted distribution is normalized to have the same sum of weights as number of
events in the data.

The classical mechanical prediction was also fitted to the data.
The probability distribution in phase space just after the entrance of the guide
is assumed to be uniform.
The same parameter definitions of $\theta$, $Z_0$, $d$ and $s$ were used,
while, for UCN loss inside the guide, two parameters $c$ and $b$ were used
instead of $\gamma$ and $\beta$.
In this case, when a UCN hits the ceiling (floor),
it is removed with a probability of $c$ ($b$) and
specularly reflected with a probability of $1-c$ ($1-b$).

\begin{figure}[tbp!]
  \includegraphics[width=85mm]{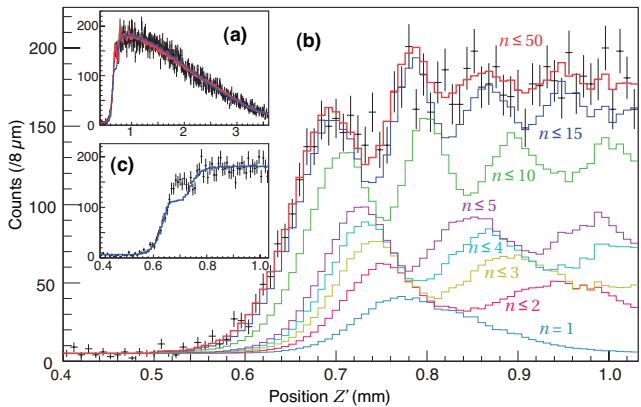}
  \caption{(color).
  (a) The observed and fitted distributions of UCNs at the pixelated detector
  for the whole region.
  Points with errors show that for the data, and the red (blue) line corresponds to
  the quantum (classical) mechanical prediction with the best fit parameters.
  (b) The cumulations of the eigenstates of the quantum mechanical prediction
  in the lower $Z'$ region.
  (c) The distributions of the classical mechanical prediction and the data
  in the lower $Z'$ region.
  \label{fig:result}}
\end{figure}

Events within the whole expected region
$0.4\;\mathrm{mm}\leq Z' \leq 3.6\;\mathrm{mm}$ are selected and used to fill
a histogram with 400 bins of width $8\;\mu\mathrm{m}$.
Eigenstates with $n\leq 50$ are used in the calculation.
The best fit parameters for quantum (classical) mechanics are
$s = 0.95\pm 0.02 \;(0.94^{+0.01}_{-0.02})$,
$Z_0 = -1.009^{+0.001}_{-0.002} \;(-0.970^{+0.002}_{-0.007}) \;\mathrm{mm}$,
$d = -0.015^{+0.004}_{-0.006} \;(-0.23^{+0.03}_{-0.02})\;\mathrm{mm}$,
$\theta = -1.254\pm0.001\;(\text{fixed to} -1.254)\;\mathrm{deg}$,
$\gamma = (9.5^{+0.7}_{-0.9}) \times 10^4 \;\mathrm{s}^{-1} \;(c=0.01\pm0.01)$ and
$\beta = 0.38^{+0.04}_{-0.03}\;(b=0.40\pm0.02)$.
The $\chi^2$ per degrees of freedom (NDF) is
$\chi^2/\mathrm{NDF} = 377.6/394 \;(439.2/395)$ 
and the corresponding p-value is 0.715 (0.062).
The distributions of the data and the predictions of the best fits are shown in
Figure \ref{fig:result}.
The form of the several modulations observed in the data is in good agreement with
the best fit prediction using quantum mechanics (Figure \ref{fig:result} (b)).
The modulation of the quantum mechanical prediction is mostly due to
eigenstates with $n \leq 15$.
This result favors the quantum mechanical prediction.
In the classical mechanical prediction (Figure\ref{fig:result} (c)),
the two rising edges at $Z'\sim 0.65\;\mathrm{mm}$ and $Z'\sim 0.75\;\mathrm{mm}$
of the distribution are due to downward- and upward-going
neutrons just before reflection by the rod.

\begin{table}[tbp!]
  \caption{Measured systematic uncertainties (with $1\sigma$).
  \label{tab:sys}}
  \begin{ruledtabular}
    \begin{tabular}{l r r}
      Source & $\Delta Z$ & $\Delta z$ \\ \hline
      $v_x$ dispersion & $1.2\;\mu\mathrm{m}$                 
          & $0.1\;\mu\mathrm{m}$\\                            
      Aberration of rod & $4.8\;\mu\mathrm{m}$                
          & $0.3\;\mu\mathrm{m}$\\                            
      Spatial resolution of detector & $3.4\;\mu\mathrm{m}$   
          & $0.2\;\mu\mathrm{m}$\\                            
      Roughness of detector surface & $10.6\;\mu\mathrm{m}$   
          & $0.6\;\mu\mathrm{m}$\\                            
      Total & $12.1\;\mu\mathrm{m}$                           
          & $0.7\;\mu\mathrm{m}$                              
    \end{tabular}
  \end{ruledtabular}
\end{table}

The estimated systematic uncertainties are summarized in Table \ref{tab:sys}.
$\Delta Z$ is the uncertainty in $Z$ and $\Delta z$ that in $z$.
These uncertainties are calculated for neutrons with $z=0$ and $v_z=0$
where the magnification power is the minimum value $16.5$.
The total uncertainty of $z$ is $\Delta z = 0.7\;\mu\mathrm{m}$.
No detector effect was found which could fake the observed modulation.
The predicted modulation is also robust against variations of fit parameters
within their uncertainties.

In conclusion, we observed the spatial distribution of gravitationally bound states
of UCNs using a novel technique.
The vertical distribution of UCNs was magnified by a cylindrical rod and detected
by a pixelated detector.
The measured UCN distribution on the pixelated detector can be derived using the Wigner
function.
The shape of several peaks of the modulation of the UCN distribution in the lower $Z'$
region is in good agreement with the quantum mechanical prediction.
This is the first observation of gravitationally bound states with submicron
resolution.


We would like to thank H.~Shimizu of Nagoya University,
V.~V.~Nesvizhevsky of Institut Laue-Langevin
and H.~Abele of the Vienna University of Technology
for their advice regarding the experiment.
We appreciate M.~Ueda of the University of Tokyo for the corroboration
on the usage of the Wigner function.
We are grateful to \mbox{T.~Brenner} of Institut Laue-Langevin for his support during
and before the experiment,
and to \mbox{S.~Sonoda} of Kyoto University for his early work on
the pixelated detector.
We also thank O.~Kirino of \mbox{Cristal Optics} Inc. for his work on the polishing
the cylindrical rod,
H.~Takeuchi of \mbox{Panasonic} Co.,\ Ltd.\ for measuring the aberration of the rod
and D.~Jeans of the University of Tokyo for proofreading the manuscript.
This work was supported by JSPS KAKENHI Grant Numbers 20340050 and 24340045 and
Grant-in-Aid for JSPS Fellows 22.1661.



\end{document}